\documentclass[preprint,aps,showpacs,showkeys,preprintnumbers,amsmath,amssymb]{revtex4}
\usepackage[dvips]{graphicx}
\usepackage{dcolumn}
\usepackage{bm}
\usepackage{rotating} 
\begin{document}
\title{Strictly finite-range potential for light and heavy nuclei}
\author{P. Salamon$^{1}$, R. G. Lovas$^{1}$, R. M. Id Betan$^{2,3}$, T. Vertse$^{1,4}$, and L. Balkay$^{5}$}
\affiliation{
$^1$MTA Institute for Nuclear Research, Debrecen, PO Box 51, H--4001, Hungary\\
$^2$Department of Physics and Chemistry (FCEIA-UNR), Avenida Pellegrini 250, S2000BTP Rosario, Argentina\\
$^3$Physics Institute of Rosario (CONICET), Bvrd. 27 de Febrero 210 bis, S2000FPA, Rosario, Argentina\\
$^4$Faculty of Informatics, University of Debrecen, PO Box 12, H--4010 Debrecen, Hungary\\
$^5$Institute of Nuclear Medicine, Medical and Health Science Center, University of Debrecen, PO Box 12, H--4010 Debrecen, Hungary
}
\date{\today}

\begin{abstract}
Strictly finite-range (SFR) potentials are exactly zero beyond their finite 
range. Single-particle energies and densities as well as $S$-matrix 
pole trajectories are studied in a few SFR potentials suited for the 
description of neutrons interacting with light and heavy nuclei. 
The SFR potentials considered are the standard cut-off Woods--Saxon (CWS) 
potentials and two potentials approaching zero smoothly: the SV potential 
introduced by Salamon and Vertse~\cite{[Sa08]} and the SS potential of 
Sahu and Sahu~\cite{[SS12]}. The parameters of these 
latter were set so that the potentials may be similar to the CWS shape. 
The range of the SV and SS potentials scales with the cube root of the mass 
number of the core like the nuclear radius itself. For light nuclei 
a single term of the SV potential (with  a single parameter) is enough 
for a good description of the neutron-nucleus interaction. The trajectories 
are compared with a bench-mark for which the starting points (belonging to 
potential depth zero) can be determined independently. Even the CWS potential 
is found to conform to this bench-mark if the range is identified with 
the cutoff radius. For the CWS potentials some trajectories show irregular 
shapes, while for the SV and SS potentials all trajectories behave 
regularly. 
\end{abstract}
\maketitle
\section{Introduction}

We call the potentials that are exactly zero beyond a certain distance
strictly finite-range (SFR) potentials. The conventional nuclear potentials 
are in principle not SFR potentials, but in practice, if the radial 
Schr\"odinger equation is solved numerically as is usual, a cutoff at a finite 
range is implied. Indeed, beyond this range $R_{\rm max}$ the numerical
solution is to be matched at a finite distance 
$r=R_{\rm match}(\ge R_{\rm max})$ with the exact solution of the 
free-particle (or of the Coulomb) problem.

For instance, the most often quoted Woods--Saxon (WS) potential goes to zero 
in infinity, but, in numerical calculations, cut-off WS (CWS) potentials 
are used invariably. A disadvantage of the CWS potential is that 
the positions of the resonance poles do depend on the cutoff distance 
\cite{[Sa08]}, which is an unphysical parameter of the calculation. 
To avoid this, a new form was introduced by Salamon and Vertse (SV) 
\cite{[Sa08]}, which contains two terms, with one range parameter for each, 
and a relative strength of the two terms. The SV potential goes to zero 
smoothly. Its  parameters can be adjusted so as to get a good fit to the WS 
shape except in the tail region, where they are necessarily different. 

There is another motivation of using SFR potentials. 
It has been observed recently by Sahu and Sahu
\cite{[SS12]} that a faster approach of the nuclear potential to zero 
improves the barrier behavior of the interaction potential between heavy ions.  
They modified  the form of the SV potential by introducing a diffuseness 
parameter $a_s$ to one of its terms. 
Here we shall refer to this potential as SS potential. 
The SS potential was found to describe the elastic scattering and 
the fusion below the Coulomb barrier with the same parameters, while a WS 
form requires two different sets for these two processes \cite{[SS12]}. 

However, the asymptotic density of the matter of nuclei is exponential, 
and the nucleon-nucleon interaction has a Yukawa tail. This physically 
substantiates the numerically untractable exponential falloff of the 
WS potential, and casts some doubt on the use of the convenient 
tails of the SV and SS potentials. In this paper we will examine 
the effect of the unphysical tail behavior of the SV potential, and further 
study the trajectories of the $S$-matrix poles. 
The SV potential is a special case of the  SS potential with $a_s=1$, 
and we extend the studies to $a_s\not=1$. In fact, for very light 
nuclei the derivative term in the SV potential can be omitted, 
and the SS form becomes identical to an SV form, which has a single 
parameter, the range $\rho_0$. 

In this work we consider nucleon potential problems. Since we disregard 
the Coulomb interaction, we can say that we deal with neutrons. 
We perform bound-state and resonance calculations, with an eye to scattering
problems, but we need no absorptive terms. We shall study the cases 
of light nuclei with mass number $A_T<20$ as well as  
nuclei with much larger $A_T$ values. Light nuclei are important 
in fusion reactions taking place in the Sun. 
The nucleon optical potential of light nuclei is an ingredient of the 
description of the reactions producing the nuclides 
used in positron emission tomography 
(PET) \footnote{The standard reactions producing the most important positron 
emitters are $^{14}$N$(p,\alpha)^{11}$C, $^{13}$C$(p,n)^{13}$N,  
$^{15}$N$(p,n)^{15}$O and  $^{18}$O$(p,n)^{18}$F.}.

\section{Functional forms of the potentials considered}

The real term of the optical potential is almost exclusively of CWS
form, and the spin-orbit part contains the derivative of 
a CWS form. 

The CWS potential can be written as 
\begin{equation}
\label{WSpot}
V^{\rm CWS}(r,R,a,R_{\rm max})=-V_0f^{\rm CWS}(r,R,a,R_{\rm max})~,
\end{equation}
with 
\begin{equation}
\label{vagottWS}
f^{\rm CWS}(r,R,a,R_{\rm max})=
\left(1+e^{\frac{r-R}{a}}\right)^{-1}~\theta(R_{\rm max}-r)~,
\end{equation}
where the Heaviside step function $\theta(x)$ is unity for positive $x$ 
and zero otherwise. The CWS form factor $f^{\rm CWS}(r,R,a,R_{\rm max})$ 
has two physical parameters, the radius $R$ and the diffuseness $a$. 
The third parameter, the cutoff radius $R_{\rm max}$, should have 
no physical significance, but, due to the jump at the finite $R_{\rm max}$, 
its derivative does not exist there, and that has implications. 

It was shown earlier~\cite{[Sa08]} that the positions of broad resonances in a 
CWS potential do depend on the value of the cutoff radius $R_{\rm max}$. 
Certain sections of the pole trajectories (mainly the starting regions) 
have been found to be sensitive to the value of 
$R_{\rm max}$~\cite{[Ra11],[Da12]}. Thus the cutoff radius 
$R_{\rm max}$ is an important, though non-physical, parameter of 
the CWS form. 
 
The SV potential~\cite{[Sa08]} recommended by two of us
instead of the CWS potential has the form \cite{[Ra11]} 
\begin{equation}
\label{SVpot}
V^{\rm SV}(r)=-V_0 f^{\rm SV}(r,c,\rho_0,\rho_1)~, 
\end{equation}
in which $V_0\ge 0$ and $f^{\rm SV}(r,c,\rho_0,\rho_1)$ is 
a linear combination of the function 
\begin{equation}
\label{distrib}
f(r,\rho)=
e^{\frac{r^2}{r^2-\rho^2}} ~ \theta(\rho-r)~,
\end{equation}
and a term containing the derivative, with respect to $r$, of the first factor, 
\begin{equation}
\label{SVder}
f^\prime(r,\rho)=-\frac{2 r \rho ^2}{(r^2-\rho^2)^2}
e^{\frac{{r^2}}{r^2-\rho^2}}~\theta(\rho-r)~. 
\end{equation}
Note that the function in Eq.~(\ref{distrib}) is a variant of the well-known 
functions of compact support, $C^\infty$, defined in the book by Bremmermann 
\cite{[Br65]} and sometimes called {\it bump functions}.
The radial factor thus contains three adjustable parameters, 
\begin{equation}
\label{newcent4}
f^{\rm SV}(r,c,\rho_0,\rho_1)=f(r,\rho_0) - c f^\prime (r,\rho_1)~,
\end{equation}
in which  $\rho_0$ and $\rho_1$ need not be the same, and, for the second term
to be attractive, the coefficient $c$ is non-negative. 
The potential $V^{\rm SV}(r)$ goes to zero smoothly, and, if $\rho_0>\rho_1$, 
it vanishes at $\rho_0$; furthermore, for $r\ge\rho_0$, it is zero, 
together with all its derivatives. Thus the SV potential has the attractive 
mathematical property that its derivative exists in the whole 
$r\in (0,\infty)$ region. A drawback is, however, that it is not analytic 
because at $\rho_0$ the Taylor series is not equal to the function. 
Nevertheless, it has turned out to be useful in quantum electrodynamics, too, 
as a compactly supported smooth regulator function \cite{[Na13]}. 

The formula of the SS potential \cite{[SS12]} is analogous to 
Eq.~(\ref{newcent4}):
\begin{equation}
\label{SSform}
f^{\rm SS}(r,c,\rho_0,\rho_1,a_s)=f(r,\rho_0) - c f^\prime (r,\rho_1,a_s)~,
\end{equation}
where 
\begin{equation}
\label{SSder}
f^\prime(r,\rho_1,a_s)=-\frac{2 r \rho_1 ^2}{(r^2-\rho_1^2)^2}
e^{\frac{a_s{r^2}}{r^2-\rho_1^2}}~~\theta(\rho_1-r)~,  
\end{equation}
with $a_s$ being the extra diffuseness parameter. 
When $a_s=1$, the SS form coincides with the SV potential~(\ref{SVpot}). 
By using $a_s \ne 1$, one naturally has more freedom in choosing the shape 
of the potential. With the usual choice  $\rho_0>\rho_1$, the range of the 
SS potential is also $\rho_0$. The SS form has the same attractive 
mathematical features as the SV potential. 

Let us return for a while to the original SV form. 
If we want the shape of the SV form to be similar to the WS shape as much as
possible, we should fit its parameters to the CWS shape $f^{\rm CWS}$. 
To this end, we can minimize 
\begin{equation}
\label{Delta}
\int_{0}^{\rho_{0}}\left[f^{\rm SV}(r,c,\rho_0,\rho_1)
-f^{\rm CWS}(r,R,a,R_{\rm max})\right]^2dr~.
\end{equation}
The integration in Eq.~(\ref{Delta}) can be performed by a quadrature of 
$m$ equidistant mesh-points $r_i=i h$ over the range of the integration, 
so that what is minimized is 
\begin{equation}
\label{Deltas}
\Delta(\rho_0,\rho_1,c)=\sum_{i=1}^{m}
\big[f^{\rm SV}(r_i,c,\rho_0,\rho_1)-f^{\rm CWS}(r_i,R,a,R_{\rm max})\big]^2~.
\end{equation}

\section{Global parameter sets for optical potentials}

In this section we construct SV potentials that approximate the real parts 
of some well-known global nucleon optical model potentials, and test 
their performance. The real parts of all global potentials are 
of CWS shape. Their geometrical shapes are generally fixed, and their 
energy dependence is restricted to the strength parameters.
The spin-orbit part for a particle with spin $s=\frac{1}{2}\hbar$ is:
\begin{equation}
\label{spinorb}
V_{\rm so}^{\rm CWS}(r,R_{\rm so},a_{\rm so},R_{\rm max})
=V_{\rm so}^{\rm CWS}h_{\rm CWS}(r,R_{\rm so},a_{\rm so},R_{\rm max})~2(
{\bf l}\cdot {\bf s})~,
\end{equation}
with a radial form 
\begin{equation}
\label{spinorbr}
h_{\rm CWS}(r,R,a,R_{\rm max})=-\frac{1}{r} f^\prime_{\rm CWS}
(r,R,a,R_{\rm max})~,
\end{equation}
in which the derivative of the central potential,
\begin{equation}
\label{derspinorb}
f^\prime_{\rm CWS}(r,R,a,R_{\rm max})=-\frac{e^{\frac{r-R}{a}}}{a
\left[1+e^{\frac{r-R}{a}}\right]^2}~\theta(R_{\rm max}-r)~,
\end{equation}
appears.

The spin-orbit term of the SV potential may be defined analogously:
\begin{equation}
\label{spinorbsv}
V_{\rm so}^{\rm SV}(r,c,\rho_0,\rho_1)=V_{\rm so}^{\rm SV}
h_{\rm SV}(r,c,\rho_0,\rho_1)~2({\bf l}\cdot{\bf s})~,
\end{equation}
with
\begin{equation}
\label{spinorbrsv}
h_{\rm SV}(r,c,\rho_0,\rho_1)=-\frac{1}{r}f^\prime_{\rm SV}(r,c,\rho_0,\rho_1)~.
\end{equation}

The mass-number dependence of the global potentials is borne generally 
by the radii such that 
$R_{\alpha}=r_{\alpha,0}A_T^{1/3}$, where $\alpha$ labels any of the 
potential terms.

Classical nucleon potential sets were given by Perey~\cite{[Pe63]} and by 
Becchetti and Greenlees~\cite{[BeGe]} long time ago, and they are relied on 
in recent studies~\cite{[Li12]} as well. A recent attempt for the 
derivation of a new $\alpha$-nucleus potential was made by Mohr and 
coworkers \cite{[Mo13]}. In this work, however we restrict 
ourselves to the Perey and Becchetti--Greenlees parameters for simplicity. 

\begin{figure}[ht]
\includegraphics*[scale=0.4, bb=0 40 750 550]{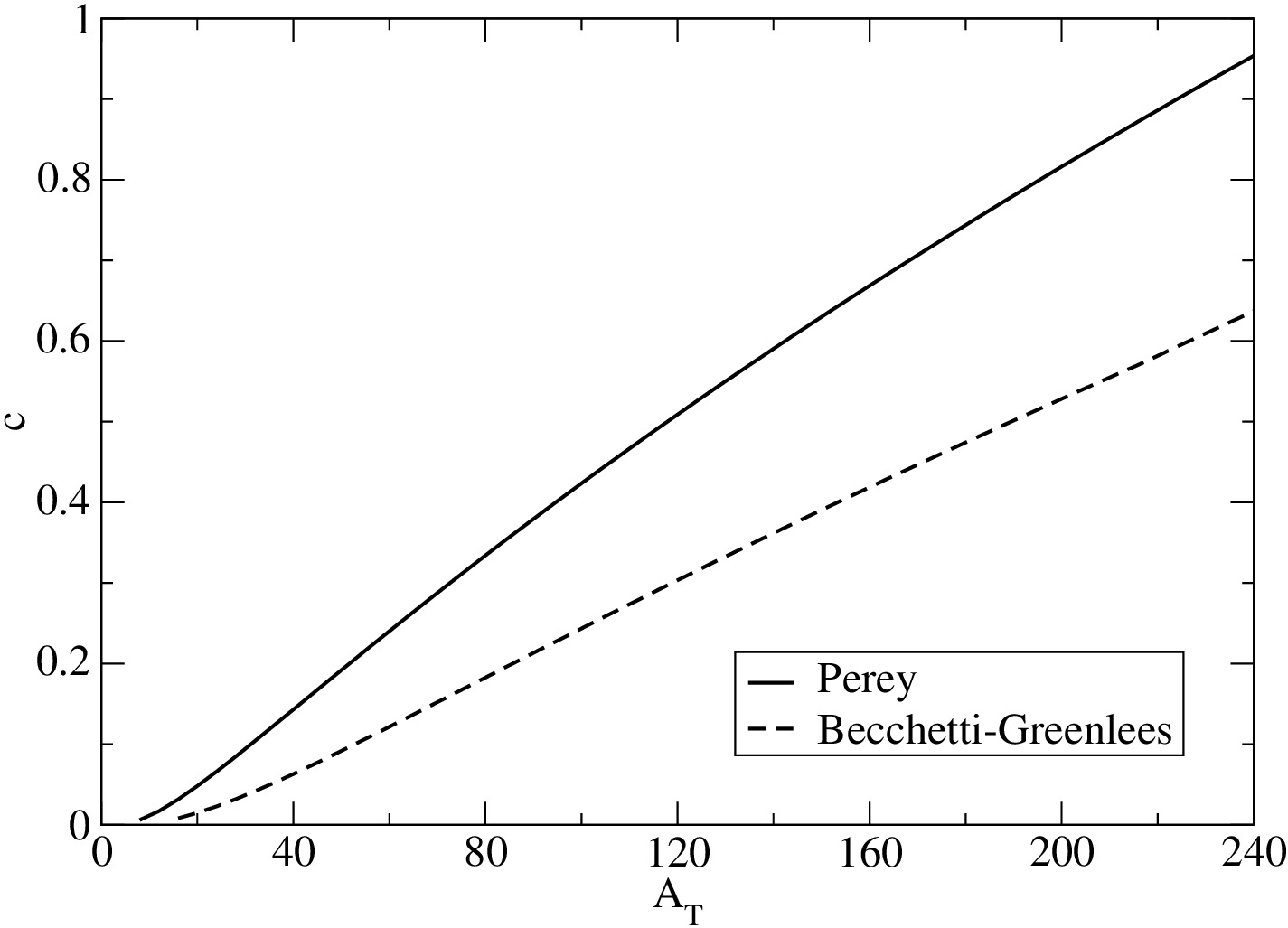}
\caption{Dependence of the mixing coefficient $c$ on the target mass-number 
$A_T$ for two global parameter sets.}
\label{c1atdep}
\end{figure}

To construct global SV potentials, we search for the minimum of the squared 
deviations in Eq.~(\ref{Delta}) as a function of the mass number $A_T$ and 
calculate the best-fit SV parameters as a function of $A_T$. For medium-heavy 
and heavy nuclei, the SV potential reproduces the CWS shape quite well, 
and its $A_T$ dependence is regular. The mixing coefficient $c$ decreases 
with decreasing $A_T$ as seen in Fig.~\ref{c1atdep}.
In the region of light nuclei, however, the best-fit SV form has a strange, 
irregular shape. We can avoid this by requiring that the derivative of 
the SV form be similar to the derivative of the WS shape: 
\begin{equation}
\label{Deltad}
\Delta(\rho_0,\rho_1,c)=\sum_{i=1}^m\left[f^{\rm SV}(r_i,c,\rho_0,\rho_1)
-f^{\rm CWS}(r_i)\right]^2+\lambda\left[{f^{{\rm SV}}}'(r_i,c,\rho_0,\rho_1)
-{f^{{\rm CWS}}}'(r_i)\right]^2~.
\end{equation}
The Lagrange multiplier $\lambda$ was determined empirically. (Here we 
suppressed the parameters of the CWS potential, which were kept fixed.) 
With a value of $\lambda=25$ fm$^2$, the fitted SV potential became reasonably
smooth and similar to the CWS shape we want to approximate.

The range $\rho_0$ of the SV potential scales with $A_T^{1/3}$, while 
the difference $\rho_0-\rho_1$ is proportional to the diffuseness $a$ 
of the CWS potential. The parameters of the
Perey potential~\cite{[Pe63]} are $r_0=1.25$ fm, and $a=0.65$ fm, and 
the best-fit SV parameters are 
$\rho_0=1.85A_T^{1/3}$ fm, $\rho_0-\rho_1=3.2a$, $c=-0.051+0.0051A_T
-3.9\times 10^{-6}A_T^2$, 
thus for small $A_T$, $c$ becomes very small.  
For the Becchetti--Greenlees~\cite{[BeGe]} geometry ($r_0=1.17$ fm and 
$a=0.75$ fm), the best-fit SV parameters relate to the CWS parameters 
very similarly, namely their values are
$\rho_0=1.86A_T^{1/3}$ fm, $\rho_0-\rho_1=2.8a$, $c=-0.055+0.003A_T-7.0
\times 10^{-7}A_T^2$.

As a light system, let us consider $^{18}$F+$n$. For the 
Perey geometry, the best-fit SV parameters are $\rho_0=5.084$ fm, 
$\rho_1=3.244$ fm, and $c=0.040$, while for the Becchetti--Greenlees
geometry, we get $\rho_0=4.957$ fm, $\rho_1=2.728$ fm, and 
$c=0.011$. This again shows that for light nuclei $c$ is practically zero, 
and it is reasonable to take $c=0$.

\begin{figure}[bht]
\includegraphics*[scale=0.4, bb=0 40 750 550]{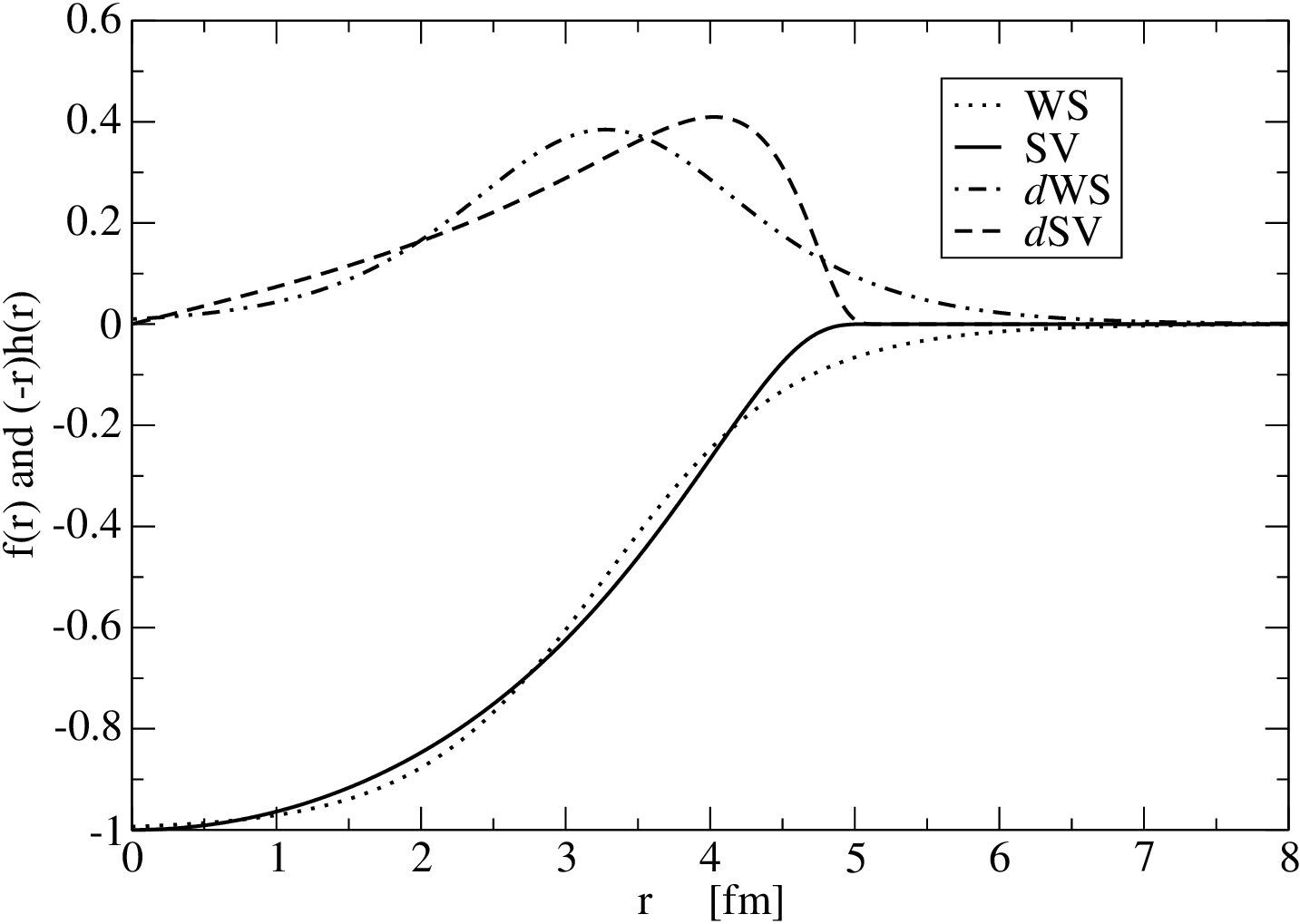}
\caption{Radial shapes of Perey's WS and the SV ($c=0$) potentials and 
their derivatives for $^{18}$F+$n$. Derivatives appear in the 
spin-orbit terms in Eqs.~(\ref{spinorbr}) and (\ref{spinorbrsv}).}
\label{a18}
\end{figure}

In Fig.~\ref{a18} we compare the shape of Perey's WS potential and its 
derivative with the SV potential (with $c=0$) and its derivative for the 
$^{18}$F+$n$ system. The WS parameters are listed in Table~\ref{compare}. The 
ratio $\rho_0/A_T^{1/3}$ is almost constant with a value of $\sim 1.6r_0$.
One can see that the radial shape of the WS potential is approximated 
reasonably well by the first term of the SV form with a single adjustable 
parameter, $\rho_0$. Now $\rho_0$ must play the role of both the radius and 
the diffuseness of the WS potential. 
Of course, the SV curves deviate most from the WS curves at large distances. 

\begin{table}[htb]
\begin{center}
\caption{Geometrical parameters of the WS and the SV potentials for $^{13}$N,
$^{15}$O and $^{18}$F. All distances are in units of fm.}
\begin{tabular}{rcccccccc}
\hline\hline
$A_T$    && $r_0\!=\!R/A_T^{1/3}$ & $R$  &&  $a$ && $\rho_0/A_T^{1/3}$ & 
$\rho_0$\\
\hline
$^{13}$N && 1.25                  & 2.94 && 0.65 &&    2.037           & 4.79 \\
$^{15}$O && 1.25                  & 3.08 && 0.65 &&    2.031           & 5.01 \\
$^{18}$F && 1.25                  & 3.28 && 0.65 &&    2.022           & 5.30 \\
\hline\hline
\end{tabular}
\label{compare}
\end{center}
\end{table}

\section{Single-particle energies for light nuclei}

It is interesting to see how the differences between the potentials influence 
the single-particle energies. In Table~\ref{spf18} we show the neutron 
single-particle energies $\epsilon_{nlj}$ calculated for the core nucleus 
$^{18}$F, with Perey's WS geometry ($V_0^{\rm CWS}=60$ MeV, $r_0=1.25$ fm, 
$a=0.65$ fm, $R_{\rm max}=15$ fm, and $V_{\rm so}^{\rm CWS}=28$ MeV). For the 
fitted SV potential we used two values for the spin-orbit strength.
In the first case 
the spin-orbit term (\ref{spinorbsv}) was used with 
$V_{\rm so}^{\rm SV}=V_{\rm so}^{\rm CWS}=28$ MeV. But, as is seen in 
Fig.~\ref{a18}, the shape of the derivative differs somewhat from that 
of the standard form. Therefore, to achieve similar spin-orbit splitting, 
in the second case we used a bit stronger ($V_{\rm so}^{\rm SV}=30$ MeV) value 
for the spin-orbit strength. 

\begin{table}[bh]
\begin{center}
\caption{$^{18}$F+$n$ single-particle energies (in MeV) in the CWS potential
and in the fitted SV potential with one central term.}
\label{spf18}
\begin{tabular}{ccccc}
\hline\hline
$i=\{n,l,j\}$&$\epsilon_i$(CWS)&\multicolumn{3}{c}{$\epsilon_i$(SV)}\\
\cline{3-5}
             &                 & $V^{\rm SV}_{\rm so}=28$ MeV 
	    && $V^{\rm SV}_{\rm so}=30$ MeV\\
\hline\hline
$0s_{1/2}$&$-38.926$& $-38.119$ && $-38.119$\\
$0p_{3/2}$&$-23.998$& $-23.568$ && $-23.611$\\
$0p_{1/2}$&$-22.067$& $-21.729$ && $-21.640$\\
$0d_{5/2}$&$-8.985$&  $-8.962$ && $-9.049$\\
$1s_{1/2}$&$-7.697$&   $-7.699$ && $-7.699$\\
$0d_{3/2}$&$-5.779$&   $-5.901$ && $-5.770$\\
\hline\hline
\end{tabular}
\end{center}
\end{table}

One can see that, with the larger spin-orbit strength, the SV energies 
are pretty close to the CWS energies. The differences are largest for the 
deepest orbits. Similar behaviors were found for the other two residual nuclei. 
In Table~\ref{core13} we present the calculated single-particle energies 
for $^{13}$N+$n$, in which the d$_{3/2}$ orbit is very close to the threshold.

We can conclude that for light nuclei the one-term SV potential is
a good phenomenological form, which reproduces the spectra obtained 
with the conventional WS potentials, although the shape of its derivative is 
somewhat different from that of the CWS potential. 

\begin{table}
\begin{center}
\caption{$^{13}$N+$n$ single-particle energies (in MeV) in the CWS potential 
and in the corresponding SV potential with one central term and 
$V^{\rm SV}_{\rm so}=30$ MeV.}
\label{core13}
\begin{tabular}{ccc}
\hline\hline
$i=\{n,l,j\}$&$\epsilon_i$ (CWS)&$\epsilon_i$ (SV)\\
\hline\hline
$0s_{1/2}$&$-35.045$&$-36.746$ \\
$0p_{3/2}$&$-18.620$&$-20.368$ \\
$0p_{1/2}$&$-16.318$&$-17.958$ \\
$0d_{5/2}$&$-3.067$&$-4.247$   \\
$1s_{1/2}$&$-3.400$&$-3.400$   \\
$0d_{3/2}$&$-0.003$&$-0.548$   \\
\hline\hline
\end{tabular}
\end{center}
\end{table}

\begin{figure}[h]
\includegraphics*[scale=0.4, bb=0 40 750 550]{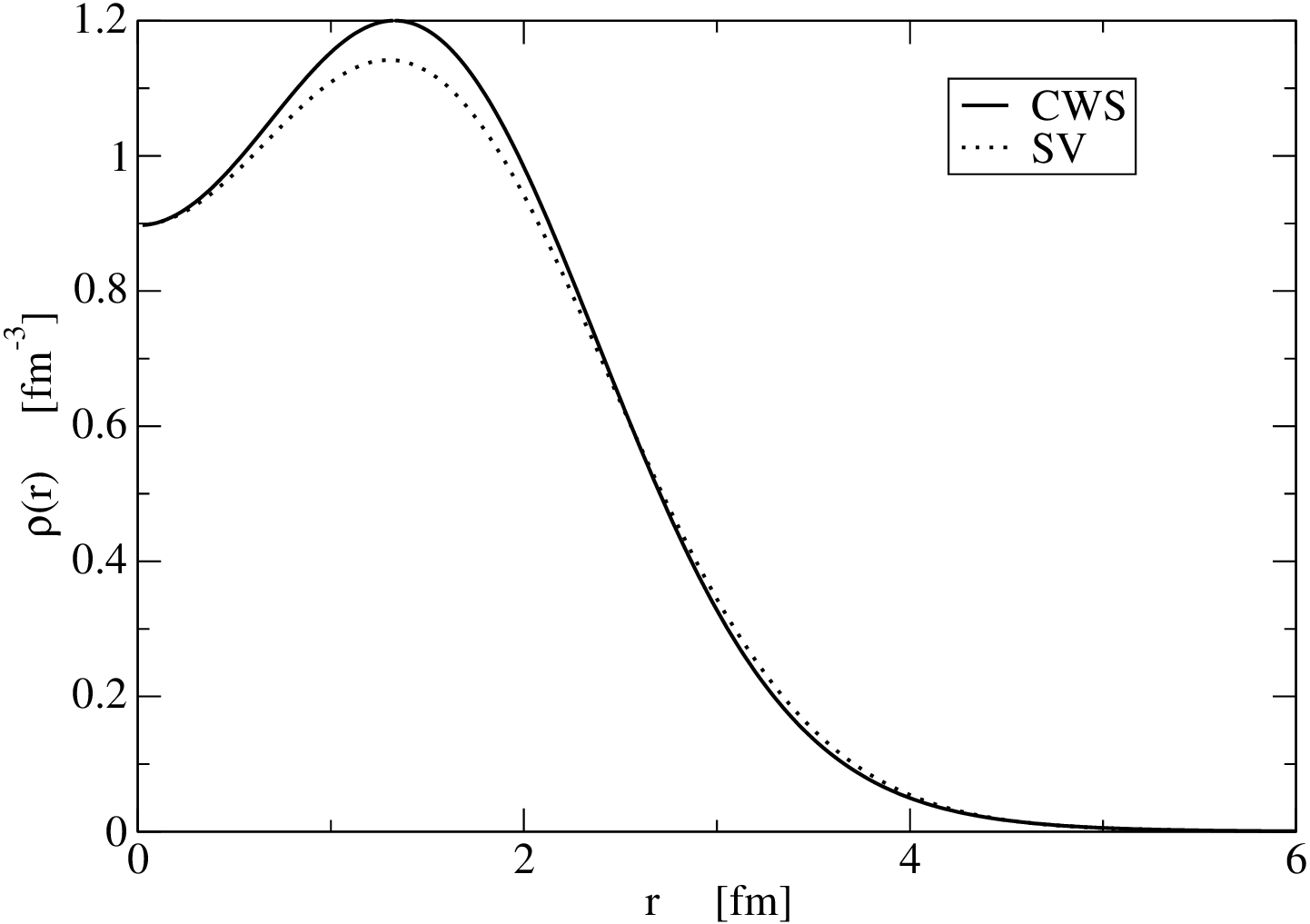}
\caption{Radial shapes of the neutron densities for the nucleus $^{18}$F 
in CWS and in SV potentials.}
\label{denshape}
\end{figure}

The wave functions produced by the two potentials are most conveniently 
compared through the neutron densities 
\begin{equation}
\label{density}
\rho(r)=\sum_i v_i^2\left[\frac{u_i(r)}{r}\right]^2~,
\end{equation}
where $i=\{n_{i},l_{i},j_{i}\}$ runs over the occupied orbits, $u_i(r)$ denotes 
the single-particle radial wave functions, and  $v_i^2$ is the occupation 
number. It is assumed that the lowest-lying orbits are fully occupied, i.e.,  
$v_i^2=2j_i+1$. In Fig. \ref{denshape} we compare the neutron 
densities calculated for the nucleus $^{18}$F in CWS and in SV potentials. 
The difference between the two densities is largest at the peak of 
the densities produced by the two deeply bound orbits, where the energies 
are deeper in the CWS potential. In the surface, where the CWS and 
SV potentials do differ appreciably, the two densities do not differ 
significantly. For $r>4$ fm, the two curves can hardly be distinguished  
because the tail of the density is mostly determined by the 
single-particle energies being close to the Fermi level, which are very similar 
in the two potentials.

\section{The CWS potential imitated by the SS form}

\begin{figure}[hbt]
\includegraphics*[scale=0.4, bb=0 40 750 550]{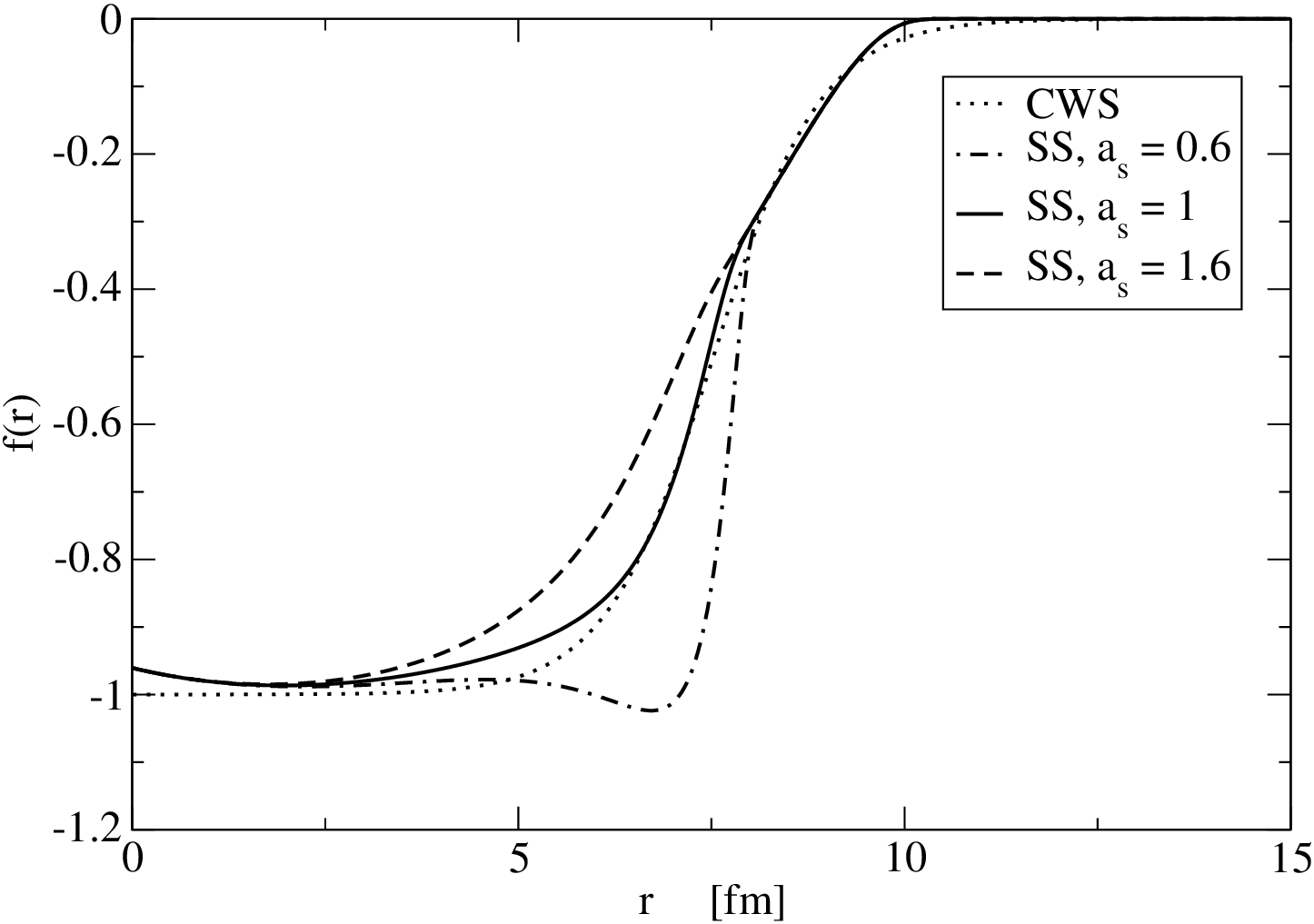}
\caption{Radial shapes of the CWS and SS potentials with different $a_s$ 
values for $^{208}$Pb+$n$.}
\label{sswspb}
\end{figure}

The SS modification only matters for heavier systems, and we consider 
$^{208}$Pb+$n$. First we show the effect of $a_s\not=1$ on a potential 
whose SV parameters $\rho_0$, $\rho_1$ and $c$ were adjusted to the CWS 
shape~\cite{[Ra11],[Da12]}. In Fig.~\ref{sswspb} we can see that $a_s>1$ 
smooths the SV potential in the region around $\rho_1$, where the SV 
curve shows a bend, while $a_s<1$ sharpens the bend, and even 
an extra minimum shows up. Such an extra minimum (a pocket) was needed 
for the description of $\alpha$ decay from Ra isotopes in Ref.~\cite{[De13]}.

\begin{figure}[htb]
\includegraphics*[scale=0.4, bb=0 40 750 550]{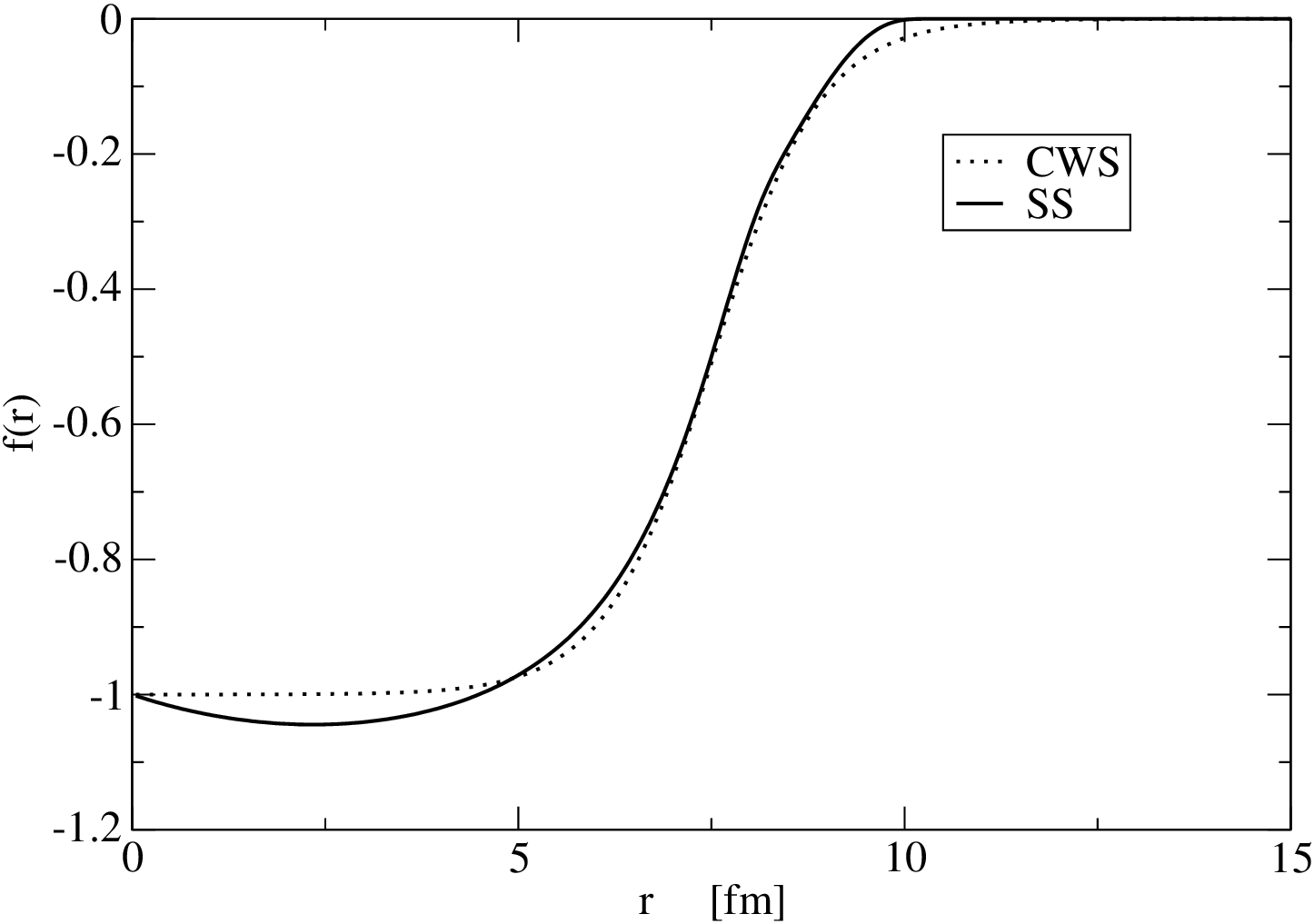}
\caption{Best-fit SS shape to the CWS shape for $^{208}$Pb+$n$. 
WS parameters: $r_0=1.27$ fm, $a=0.7$ fm. SS parameters: 
$\rho_0=10.75$ fm, $\rho_1=8.94$ fm, $c=1.528$, $a_s=1.4$.}
\label{fittedss}
\end{figure}

To determine the SS form that approximates the CWS potential best, we should 
fit all the four parameters of the SS potential simultaneously.
We  minimized the function
\begin{equation}
\label{Deltad1}
\Delta(\rho_0,\rho_1,a_s,c)
=\sum_{i=1}^m\left[f^{\rm SS}(r_i,c,\rho_0,\rho_1,a_s)
-f^{\rm CWS}(r_i)\right]^2
+\lambda\left[{f^{\rm SS}}'(r_i,c,\rho_0,\rho_1,a_s)
-{f^{\rm CWS}}'(r_i)\right]^2~,
\end{equation} 
with  $\lambda=25$ fm$^2$. The two potentials are shown in Fig.~\ref{fittedss}, and 
the parameters are given in the caption. 
The agreement is remarkable in spite of the SS potential having a minimum.

\section{Pole trajectories in SFR potentials}

Having indicated some practical aspects of using SFR potentials in 
nuclear problems, we now discuss the problem of pole trajectories. 
We remind the reader that it is the pole trajectories, especially 
in the region of broad resonances, that make the use of truncated 
potentials dangerous. Pole trajectories can be labeled conveniently 
by $n$, the number of nodes of the wave function defined where the 
pole belongs to a bound (or anti-bound) state. However, the trajectories can be 
found more easily at the other extreme, where the potential strength is nearly 
zero (at the ``starting point''). Here the states are resonances 
with complex radial wave functions, whose real as well as imaginary 
parts have infinite numbers of zeros. Orbits with low $n$ values are 
important in nuclear structure calculations and in low-energy nucleon 
scattering. In heavy-ion reactions larger $n$ values occur. 
In the present work we restrict ourselves to the s-wave case. Analytical 
results are available for the square-well potential in the work of 
Nussenzweig \cite{[Nu59]} as was discussed by some of us recently 
\cite{[Da12]}. Since, however, we are concerned with less special potentials, 
which cannot be treated analytically, we re-consider approximate analytical 
formulae for the starting points of the trajectories given in the literature.  
We are interested in where these are valid and how they can be treated 
numerically. 

\subsection{Formulae for the starting points}

The $l=0$ states in the SFR potential
\begin{equation}
\label{newpot}
V(r)=V_0~\theta({\cal R}-r)[({\cal R}-r)^\sigma +\ldots]~
\end{equation}
are discussed by R. G. Newton in his book~\cite{[Ne82]} [see Eq.~(12.98) 
on p.~361 there]. Here $\sigma>0$, $\theta(x)$ denotes the 
Heaviside step function, and the square bracket contains a truncated 
expansion in terms of ${\cal R}-r$. In Eq.~(12.102) on p. 362 Newton gives 
the real and imaginary parts of the starting point $k_n=k_n^R -{\rm i} k_n^I$
of the trajectory of the $n$th pole of the $S$-matrix as follows:
\begin{equation}
\label{rek}
k_n^R= \frac{n\pi}{\cal R}+O(1)~,
\end{equation}
and
\begin{equation}
\label{imk}
k_n^I= \frac{\sigma+2}{2{\cal R}}\ln (n) +O(1)~.
\end{equation}
The starting point of the pole trajectory is in the fourth quadrant of the 
$k$-plane, and, by definition, it belongs to $V_0=0$. Equations~(\ref{rek}), 
(\ref{imk}) are especially useful for large $n$ values, where the $O(1)$ 
terms in the equations can be neglected, but it is interesting to see 
how they are fulfilled for lower $n$.
Eq.~(\ref{rek}) depends linearly on $n$ with a slope
\begin{equation}
\label{exslope}
A_1=\frac{\pi}{\cal R}~.
\end{equation}

Regge pointed out \cite{[Re58]} that a relation similar to Eq.~(\ref{rek}) is 
valid for the moduli of the starting wave number values: 
\begin{equation}
\label{absk}
|k_n|= \frac{n\pi}{\cal R}+O(1)=A_1 n+O(1)~.
\end{equation}

\subsection{Test with Newton's potential}

For a potential of the form of (\ref{newpot}), the asymptotic expressions 
(\ref{rek}),(\ref{imk}) and (\ref{absk}) offer convenient tests of our 
numerical procedure for very large $n$ values.
Inaccuracies may come from approximating $V_0=0$ by a small finite value, 
from truncation errors in the numerical integration of the differential 
equation, and from rounding errors throughout the numerical calculations.
We reduced the rounding errors by using extended precision floating-point 
arithmetics. We used Ixaru's method \cite{[Ix84]} for the numerical integration 
of the radial equation, and we calculated the position of the pole of the 
$S$-matrix using the computer code ANTI \cite{[anti]}. 

We chose a potential of the form of Eq.~(\ref{newpot}) with $\sigma=1$:
\begin{equation}
\label{ournewpot}
V(r)=-V_0~\theta({\cal R}-r)({\cal R}-r)~,
\end{equation}
which is attractive if $V_0>0$, and chose $V_0=0.005$ MeV 
and ${\cal R}=10$ fm.

\begin{figure}[h]
\includegraphics*[scale=0.4, bb=0 40 750 550]{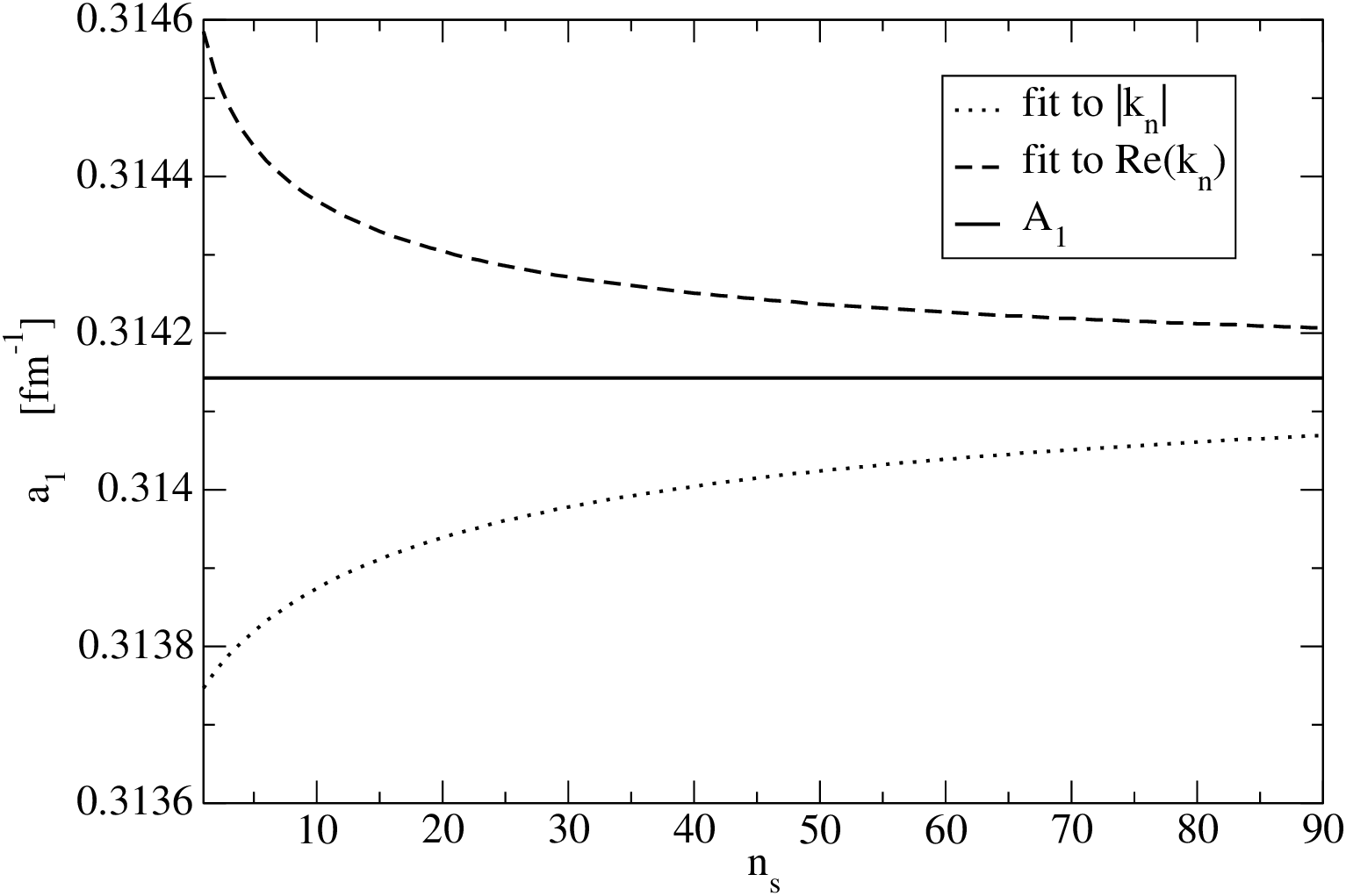}
\caption{Dependence of the slope of the fitted line on the lower cut value
of the node number $n_s$ for a potential in Eq.~(\ref{ournewpot}) 
with a range of ${\cal R}=10$ fm.}
\label{c1}
\end{figure}

We calculated the starting values $k_n$ for the $n=1,\ldots,98$ trajectories, 
and fitted the $k_n^R$ values by a first order polynomial of $n$, i.e.,
\begin{equation}
\label{firstpol}
y(n)=a_0+a_1 n~.
\end{equation}
Since in Eq.~(\ref{rek}) we have an unknown $O(1)$ term (the actual value of 
this is reflected by $a_0$), 
we applied a lower cut value $n_s$ in our data and performed the fitting for a 
number of $n\in \{n_s,n_s+1,\ldots,n_u\}$ with $n_u=98$ fixed and $n_s$ varied. 
We can thus estimate the value of $a_1$ for each $n_s$ and compare it with 
$A_1=\pi/{\cal R}=0.31416$ fm$^{-1}$ obtained from Eq.~(\ref{exslope}). In 
Fig.~\ref{c1} the ordinate shows the deduced slope, with the horizontal line 
$A_1=\pi/10$ fm$^{-1}$, to which the fitted values of $a_1$ should converge for 
large $n_s$. 
The dashed line connects the $a_1$ values resulting from the fit to 
$k_{n_s}^R$. 
It is seen that the estimate for the 
range has 3 accurate digits even for $n_s=1$.

To check the validity of Eq.~(\ref{absk}), we fitted a linear function to
the moduli of the starting wave number values calculated, and 
followed a procedure similar to that for $k^R_n$.  
The dotted line in Fig.~\ref{c1} shows the slopes obtained as a function of 
$n_s$. Now the fitted slope $a_1$ approaches the horizontal line from below 
and yields an estimate of similar accuracy.  
The results of these tests show that the small final value of $V_0$ 
we use provides a reasonable estimate for the starting value of the 
pole trajectory. 

\begin{figure}[h]
\includegraphics*[scale=0.4, bb=0 40 750 550]{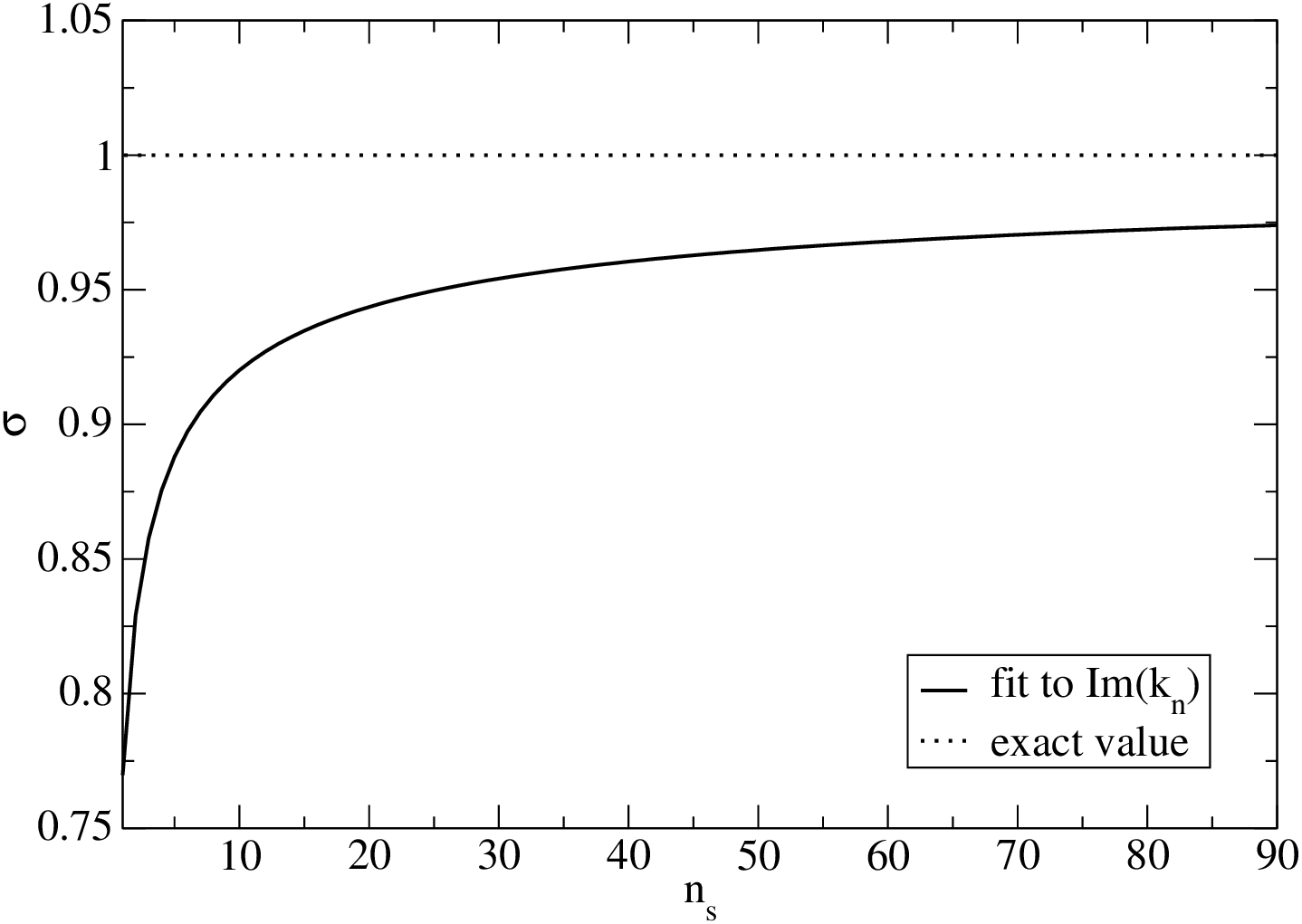}
\caption{Convergence of the fitted $\sigma$ to the exact value (dotted line) 
obtained by using the lower cut value of the node number $n_s$ for a potential 
of Eq.~(\ref{ournewpot}).}
\label{sigmans}
\end{figure}

To check Eq.~(\ref{imk}) for the imaginary part of $k_n$, we introduce 
the variable $x=\ln (n)$ and fit $k_n^I=a_1x+a_0$ 
for the same sets of $n=n_s,\ldots,98$ points, with $n_s=1,\ldots,97$.
From the slope $a_1$ obtained, we can calculate $\sigma=2a_1{\cal R}-2$ 
as a function of $n_s$ using the actual value of ${\cal R}$.
Figure~\ref{sigmans} shows that this $\sigma$ converges 
to 1 as it should, but rather slowly. 

\subsection{Cut--off Woods-Saxon form}

The trajectories of the $S$-matrix poles were calculated for two SFR potentials
for a heavy nucleus $^{208}$Pb in Refs.~\cite{[Ra11],[Da12]}.
Certain features found in Ref.~\cite{[Da12]} indicate that the 
relationship~(\ref{rek}) might hold for the CWS and even for the 
SV potentials.

The asymptotic behavior of the CWS potential for $r<R_{\rm max}$ may be 
approximated by a Taylor series around $r={\cal R}=R_{\rm max}$ cut after the 
first term:
\begin{equation}
\label{taylorWS}
-V_0f^{\rm CWS}(r,R,a,R_{\rm max})\approx {D}+(R_{\rm max}-r)\frac{D}{a}~,
\end{equation}
where $D=-V_0e^{(R-R_{\rm max})/a}$.
The second term corresponds to a $\sigma=1$ version of Newton's potential
studied before, but now we have an additional first term, which does not depend 
on $r$. Thus not even an approximation to the CWS potential has exactly the 
form of Eq.~(\ref{newpot}). But, with the usual choice of 
$R_{\rm max}\ge R+6 a$, the value of the constant $|D|\le 0.0025\times V_0$, 
thus the first term is not very large.

Since for a heavy nucleus, a crucial difference has been observed between 
the pole trajectories of the continuous SV potential and the discontinuous 
CWS potential~\cite{[Ra11],[Da12]}, here we extend these calculations 
to light nuclei and to the SS potential. 

\begin{figure}[h]
\includegraphics*[scale=0.4, bb=0 30 750 550]{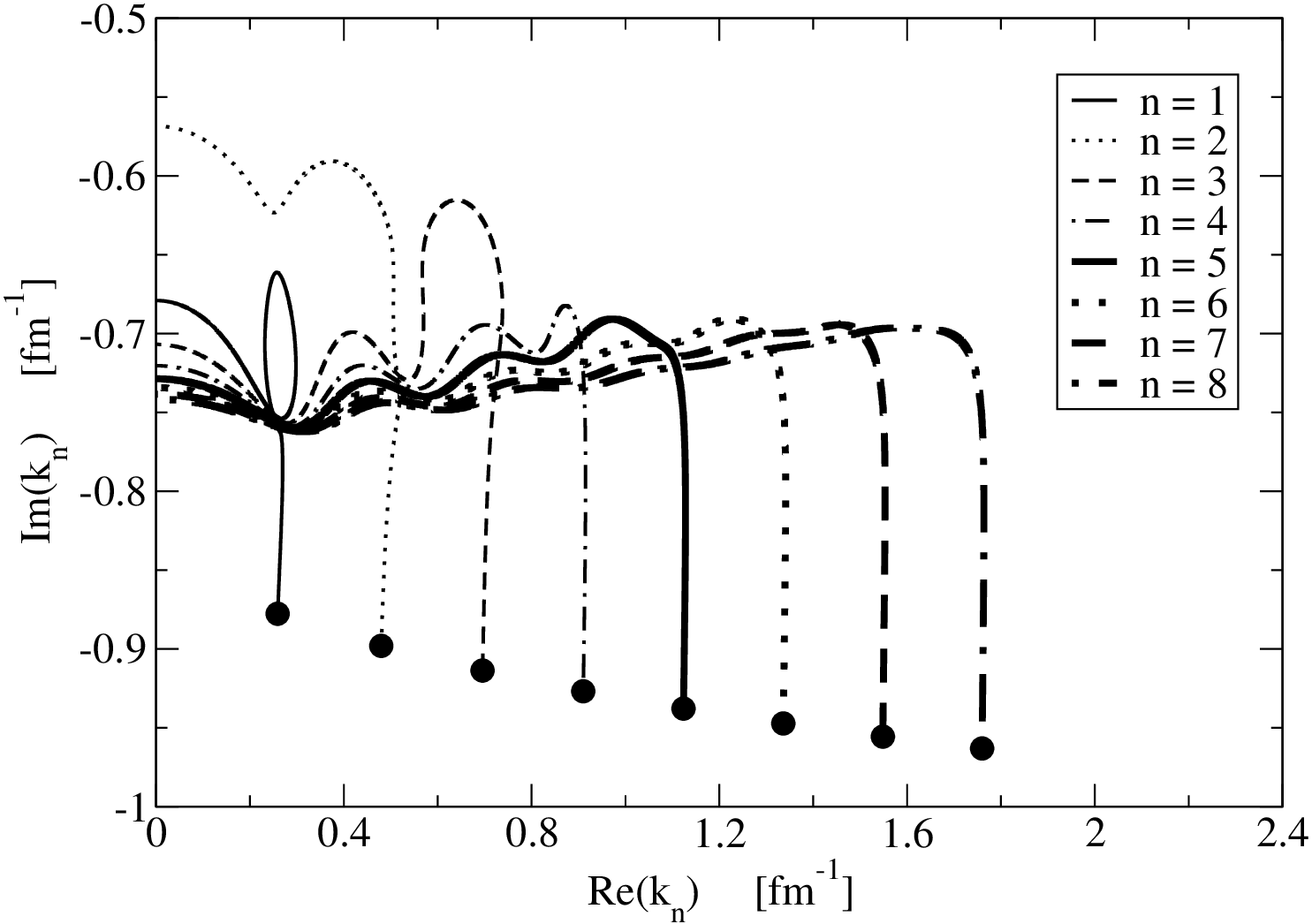}
\caption{Pole trajectories for a CWS potential with $R_{\rm max}=15$ fm for 
$l=0$ and $n=1,\ldots,8$ for $^{18}$F. The full circles denote the starting 
points of the trajectories with $V_0=0.005$ MeV.}
\label{wstraj}
\end{figure}

\begin{figure}[h]
\includegraphics*[scale=0.4, bb=0 40 750 550]{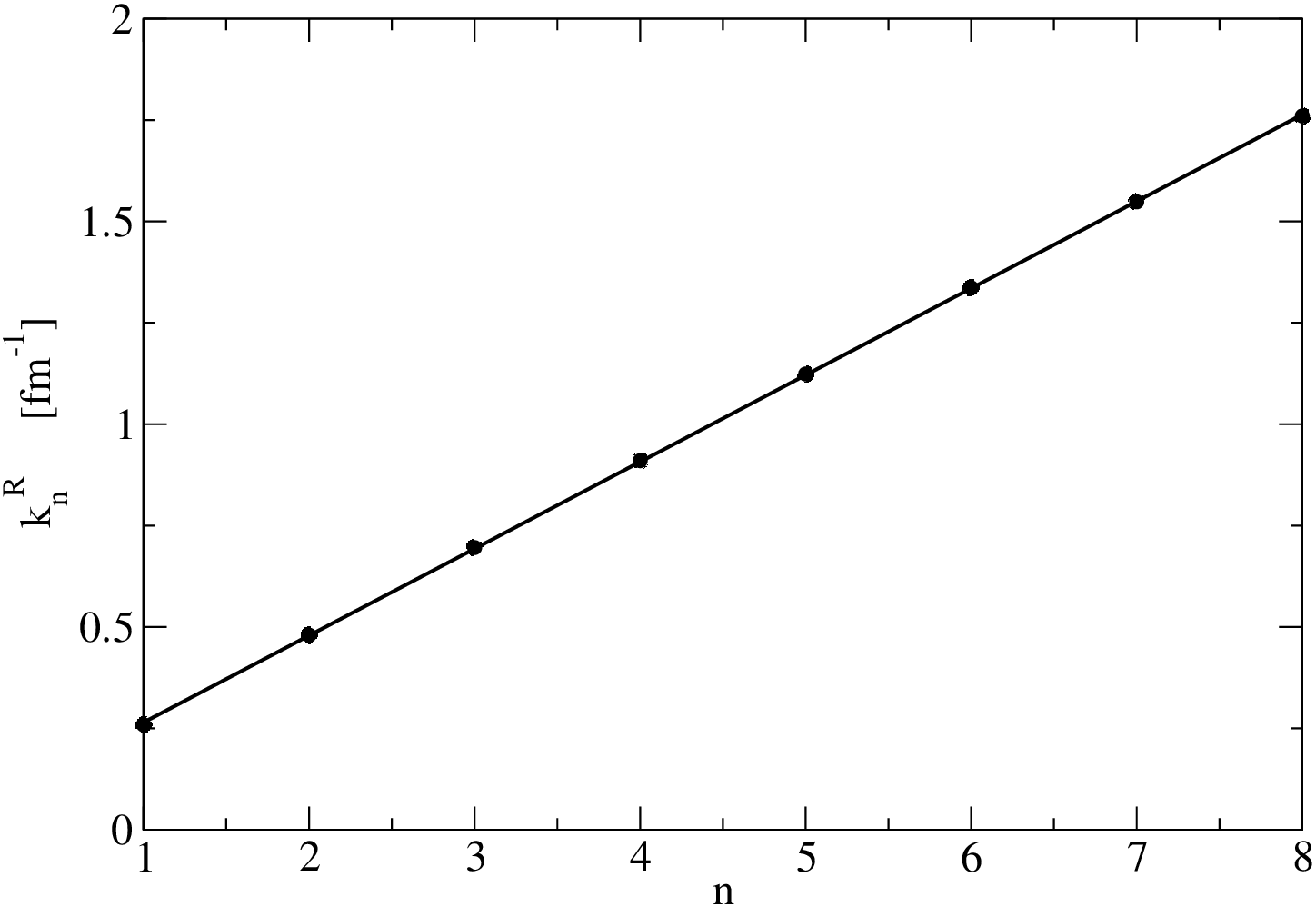}
\caption{The line is the linear function fitted to the $k_n^R$ values (dots) 
of the pole trajectories with node numbers $n=1,\ldots,8$ for a CWS potential 
for $^{18}$F. These values correspond to the abscissae of the full circles in
Fig. \ref{wstraj}. The fit results in a range ${\cal R}=14.67$ fm.}
\label{f18cws}
\end{figure}

For $^{208}$Pb, it has been found~\cite{[Da12]} that the starting points 
of the $l=0$ resonant trajectories follow Newton's rule in Eq.~(\ref{rek})
approximately if the $n$ value is not very small even though the asymptotic 
behavior of the potential~(\ref{taylorWS}) differs slightly from 
Eq.~(\ref{newpot}). Figure~\ref{wstraj} shows the trajectories of a few poles 
of the $^{18}$F+$n$ system in the CWS well with parameters $r_0=1.25$ fm,  
$a=0.65$ fm, and $R_{\rm max}=15$ fm. The results are similar to those for 
$^{208}$Pb even in that there is a loop in the $n=1$ trajectory but nowhere 
else. Figure~\ref{f18cws} shows the straight line fitted to $k_n^R$ for 
node numbers $n=1,\ldots,8$. From its slope Eq.~(\ref{exslope}) predicts 
${\cal R}=14.67$ fm, which agrees reasonably well with the cutoff radius 
used, $R_{\rm max}=15$ fm ($|D|=1.4 \times 10^{-8}V_0$). 

\begin{figure}[h]
\includegraphics*[scale=0.4, bb=0 40 750 550]{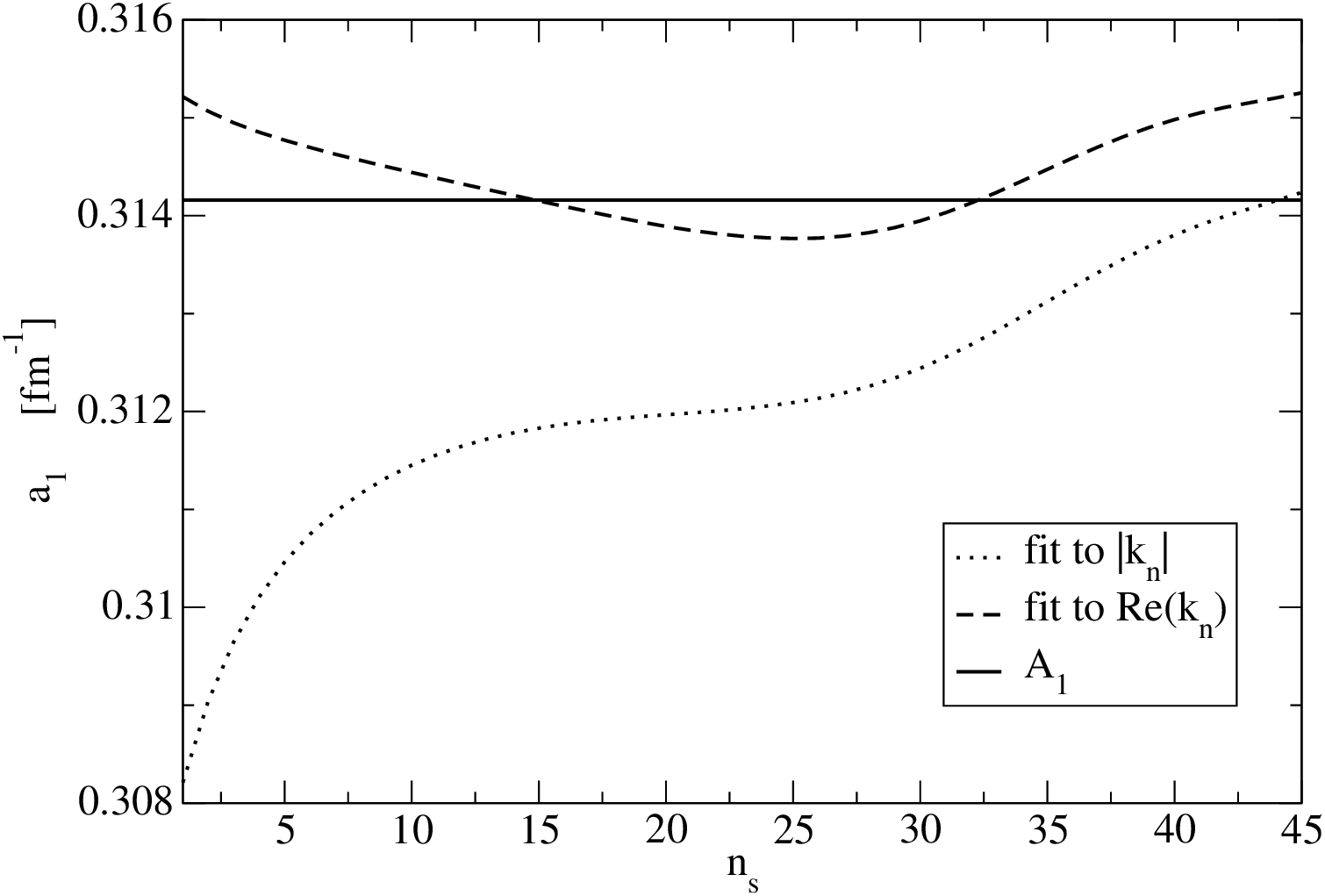}
\caption{Dependence of the slope of the fitted line on the lower cut value
of the node number $n_s$ ($n_s=1,\dots,n_u-1$, and $n_u=48$) for a CWS potential with $R_{\rm max}=10$ fm.}
\label{c1ws}
\end{figure}

We studied the behavior of the trajectories further by setting the cutoff 
radius shorter, $R_{\rm max}=10$ fm ($|D|=3.2\times 10^{-5}V_0$).  
In Fig.~\ref{c1ws} we examine the validity of Eqs.~(\ref{rek}) and (\ref{absk}) 
for the CWS potential by a test similar to that shown in Fig.~\ref{c1}. 
Now the two curves do not converge smoothly into a constant. 
The agreement of the slope $a_1$ with the exact value 
is reduced to 2 decimal digits, and, as a function of $n$, it oscillates around 
$\pi/R_{\rm max}$. Thus we can still state that Eqs.~(\ref{rek}) and 
(\ref{absk}) are approximately satisfied by a CWS potential as well.
The relationship for the imaginary part, Eq.~(\ref{imk}), however, is not 
satisfied at all. There is no region where the deduced 
$\sigma$ would be more or less constant. 
It looks that Eq.~(\ref{taylorWS}) is too approximate to cause Eq.~(\ref{imk}) 
to be fulfilled. 

\begin{figure}[b]
\includegraphics*[scale=0.4, bb=0 30 750 550]{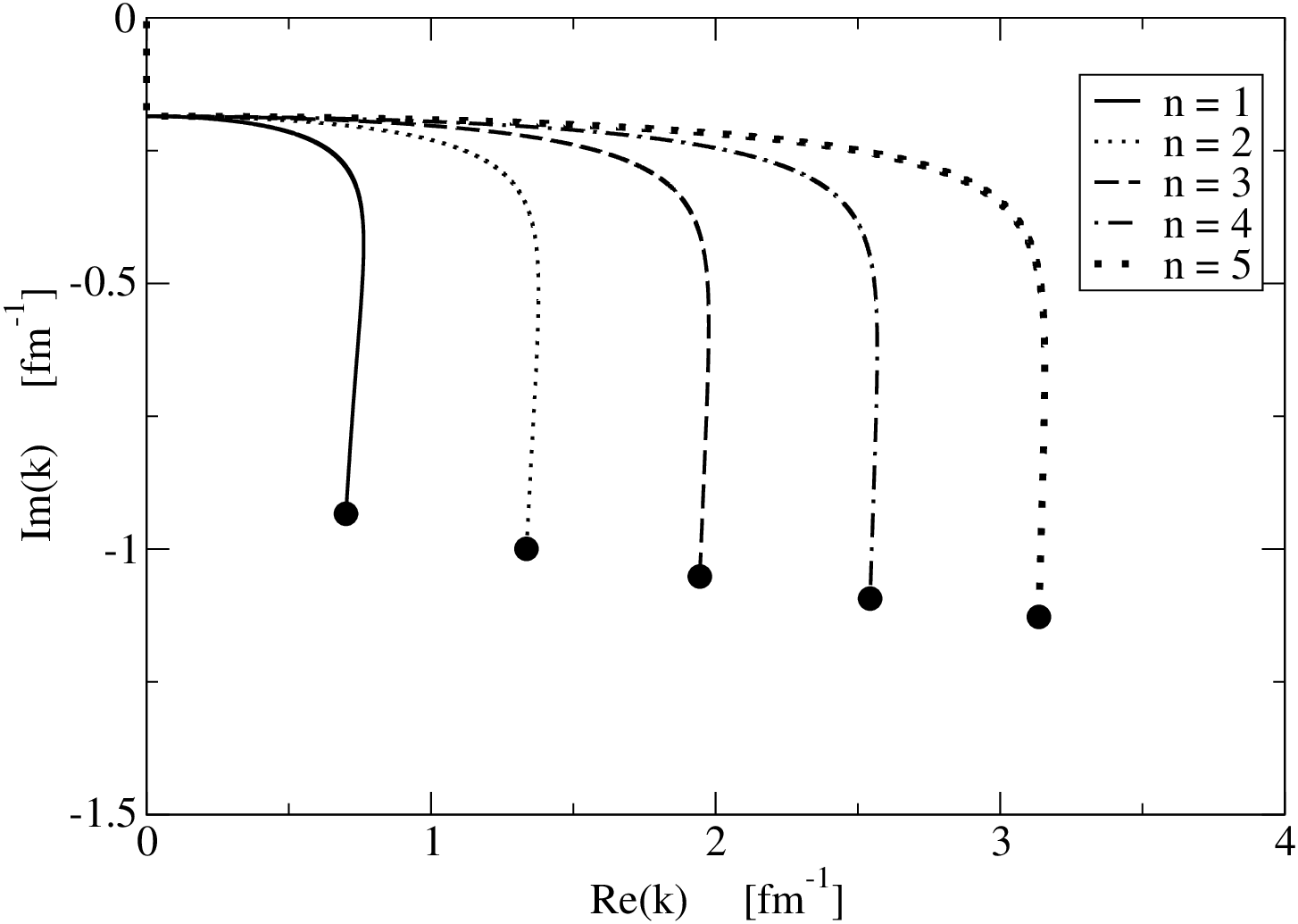}
\caption{Pole trajectories for a SV potential with $\rho_0=5.3$ fm for $l=0$ 
and $n=1,\ldots,5$ for $^{18}$F. The full circles denote the $k_n$ points 
calculated with $V_0=0.005$ MeV.}
\label{svtraj}
\end{figure}

\subsection{Pole trajectories in SV and in SS potentials}
 
\begin{figure}[h]
\includegraphics*[scale=0.4, bb=0 30 750 550]{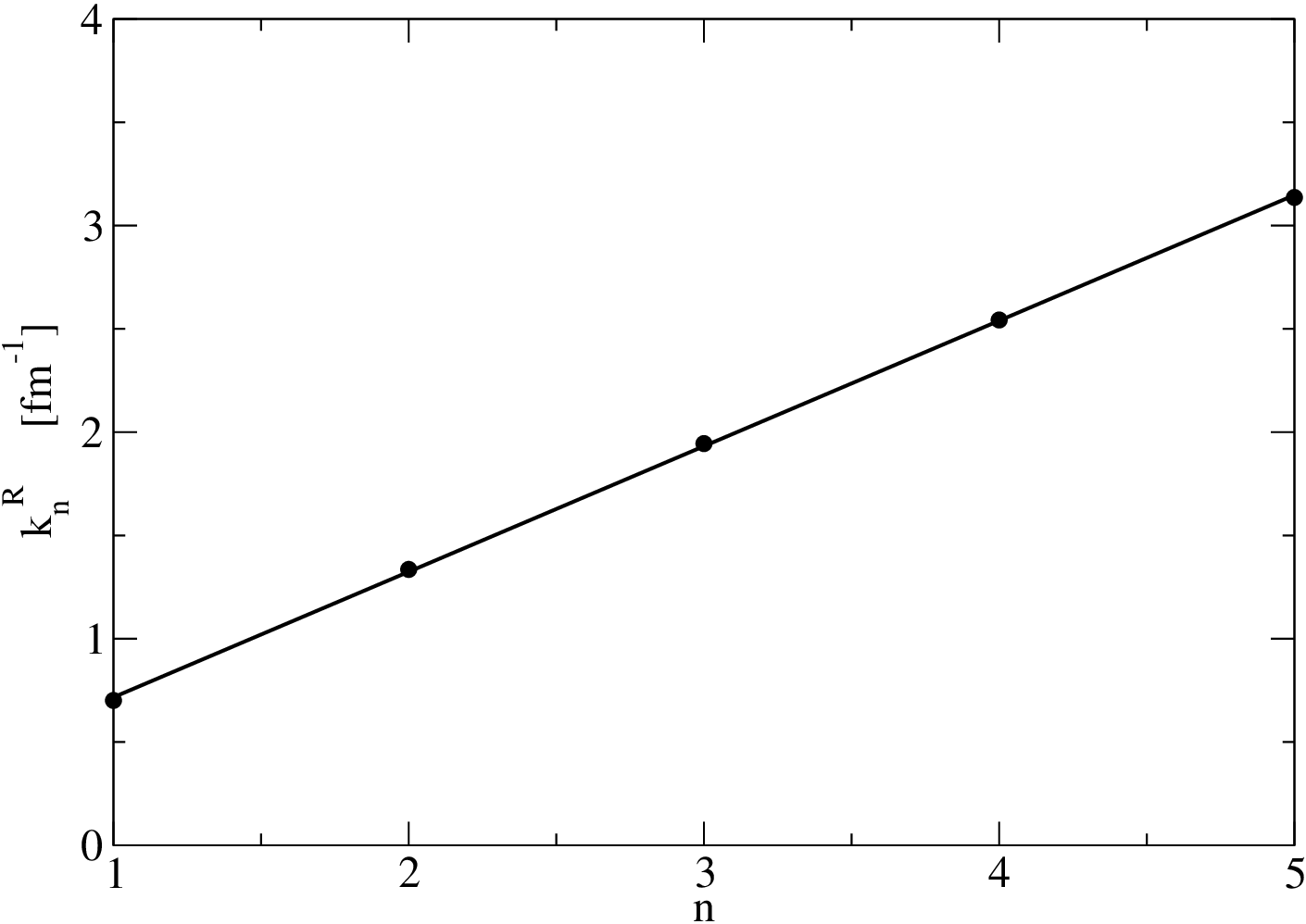}
\caption{Fit  to the $k_n^R$ values of the full circles in Fig.~\ref{svtraj}, 
with node numbers $n=1,\ldots,5$ for a single-term SV potential for $^{18}$F. 
The range deduced from the slope $a_1$ is ${\cal R}=5.17$ fm.}
\label{sv18slope}
\end{figure}

The pole trajectories for the SV potential behave absolutely regularly, with 
no loops and ripples (Fig.~\ref{svtraj}), in contrast to the CWS potential. 
The starting values $k_n^R$ can be fitted very well by a straight line as
seen in Fig.~\ref{sv18slope}. From its slope and Eq.~(\ref{rek}) one
can derive ${\cal R}=5.17$ fm, which is just a bit less than the value of the 
range parameter $\rho_0=5.3$ fm. Similar behavior was found before for 
$^{208}$Pb in Ref.~\cite{[Da12]}.  We conclude that the relation in 
Eq.~(\ref{rek}) is fulfilled approximately for SV and SS potentials in spite 
of their asymptotic behavior being different from Eq.~(\ref{newpot}). Thus 
Eq.~(\ref{rek}) is still useful for estimating the pole positions.
Remember that the SV and SS potentials the Taylor expansion at 
$\rho_0$ is not equal to the function, because all derivatives are zero at that point.

For two-term SV potentials ($c\ne0$), the starting values of the pole 
trajectories were studied in Ref.~\cite{[Ra11]} for $^{16}$O and 
for $^{208}$Pb \footnote{In Ref.~\cite{[Ra11]} it was conjectured that, for 
low node numbers, the $k_n^R-k_{n-1}^R$ is determined 
by ${1\over 2}(\rho_0+\rho_1)$. Later it turned out that this result was 
just an accident. The starting point depends only on
$\rho_0$, where the potential vanishes.}.
Now these studies may be extended to the SS potentials of various $a_s$. 
If the SS potential obeyed Newton's relation (\ref{newpot}), 
the starting regions should be independent of $a_s$ and should coincide 
with the SV trajectory.
Since, however, Eq.~(\ref{newpot}) does not hold even for the SV potential, 
we expect a dependence. 

\begin{figure}[h]
\includegraphics*[scale=0.4, bb=0 30 750 550]{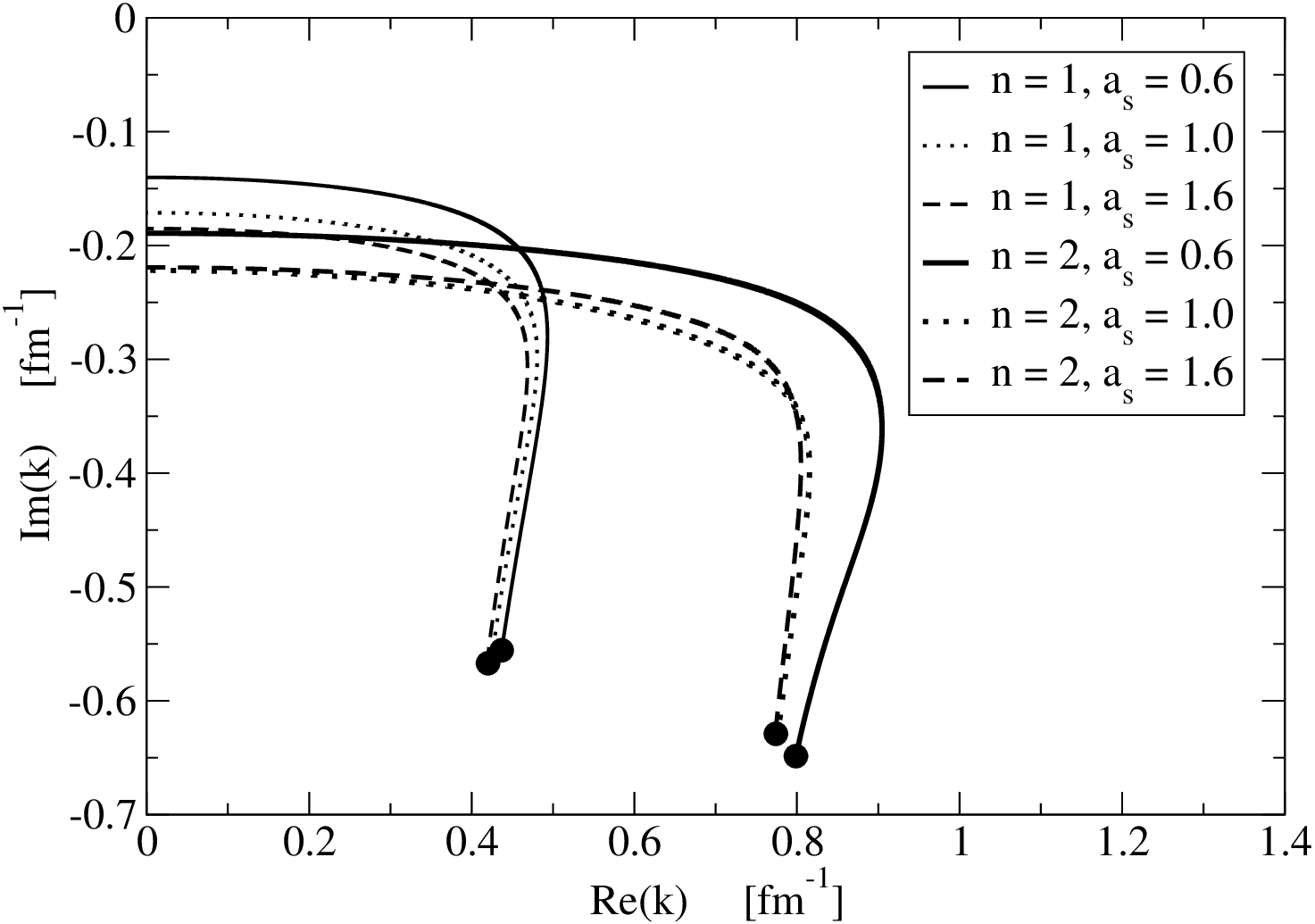}
\caption{Pole trajectories in SS potentials with different $a_s$ values. The 
value $a_s=1.0$ corresponds to the SV potential. The full circles denote 
the $k_n$ values calculated with $V_0=0.005$ MeV.}
\label{sstraj}
\end{figure}

We consider a heavy core, where the derivative term is important: the case of 
$^{208}$Pb+$n$. We choose $l=0$, analyze the SV potential that 
approximates the CWS potential of parameters $R=7.525$ fm and $a=0.7$ fm 
($\rho_0=10.963$ fm, $\rho_1=8.328$ fm, and $c=0.997$), 
and repeat the calculation for SS potentials of $a_s=0.6$ and 1.6 
(Fig.~\ref{sstraj}). One can see that the three curves belonging to the same 
$n$ do not coincide, and nor do their starting points, but they slightly 
depend on $a_s$. This weak dependence may be 
attributed to departures from Eqs.~(\ref{rek}) and (\ref{imk}) for 
low $n$.

\begin{figure}[h]
\includegraphics*[scale=0.4, bb=0 30 750 550]{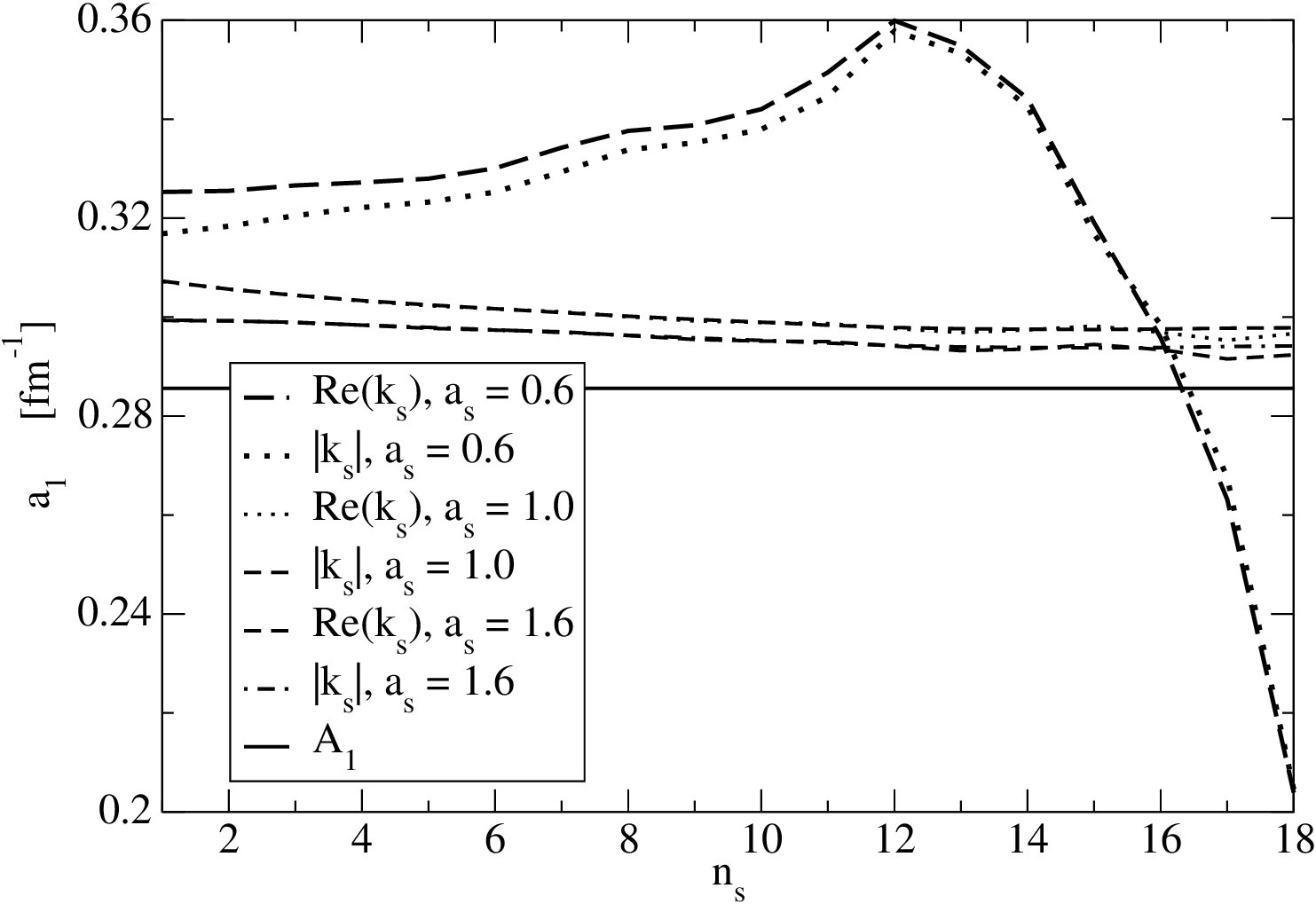}
\caption{Slope $a_1$ of the straight line fitted to the starting $k_n$ 
values ($n_s=1,\dots,n_u-1$, and $~n_u=20$) for SS potentials of different $a_s$, with $a_s=1.0$ belonging 
to the SV potential; $A_1=\pi/11$ fm$^{-1}$. Slopes belonging to $a_s=1.0$ and $a_s=1.6$ are hardly distinguishable in the given scale.}
\label{ssa1}
\end{figure}

We calculated the starting $k_n$ values for the best-fit SS shape to the same 
CWS shape for $^{208}$Pb+$n$. 
(WS parameters: $r_0=1.27$ fm, $a=0.7$ fm. SS parameters: 
$\rho_0=10.75$ fm, $\rho_1=8.94$ fm, $c=1.528$, $a_s=1.4$).
Although we know that the SS potential does not follow Newton's
form [Eq.~(\ref{newpot})], we can still fit our $k_n$ values by first-order 
polynomials of the variable $n$ and $\ln(n)$, respectively, to check 
the validity of Eqs.~(\ref{rek}), (\ref{absk}) and (\ref{imk}).
Equations~(\ref{rek}) and (\ref{absk}) seem to be valid approximately in the 
$n$-range shown in Fig.~\ref{ssa1} for $a_s\ge 1$. For $\sigma=0.6$, which
produced a pocket in Fig.~\ref{sswspb}, the relation breaks down beyond 
$n_s\approx 12$.

Test calculations show that the $k_n^I$ values weakly depend on $a_s$, and the
$\sigma$, defined by Eq.~(\ref{imk}), does not seem to converge. That is 
not surprising as neither the SV nor the SS potential satisfies 
Eq.~(\ref{newpot}). Just as for the SV potential, the $k_n^I$ values show 
an almost linear slow increase with $n$. This offers practical recipes 
for finding suitable starting values in searches for $S$-matrix poles.

\section{Conclusion}

The conventional nuclear potentials  
do not tend to zero at finite distances, but are set to zero artificially. 
Consequently, they have unpleasant mathematical and numerical properties, 
which cause appreciable errors in broad resonances. Their SFR substitutes 
have pleasant mathematical and 
numerical properties, but their tails are unphysical. Here we 
examined the properties of a family of SFR potentials related to 
the WS potential, with an emphasis on the effect of the tail and on 
the pole trajectories belonging to broad resonances. 

We concentrated on the SV potential, which consists of a term 
$\exp[(r^2/(r^2-\rho_0^2)]$ 
($r<\rho_0$) and a term like the derivative 
of that but with a different parameter $\rho_1$ ($\le\rho_0$). 
We constructed parameters that fit 
the real parts of the global Perey--Perey and Becchetti--Greenlees 
optical potentials best. 
The best-fit range $\rho_0$ of the SV potential is found to scale by 
$A_T^{1/3}$ for both geometries, and 
the difference of the two ranges, $\rho_0-\rho_1$, is positive and 
it is three to four times of the diffuseness of the WS potential. 
The admixture of the derivative term tends to zero with decreasing mass number. 

In fact, it was found that, for light nuclei, the phenomenological neutron 
potential can be approximated reasonably well by a single-term SV potential, 
and the single-particle energies and densities calculated in the cut-off
WS potential are also reproduced. In this case the form factor of the potential 
has a single parameter, its range $\rho_0$. The tail of the density is pretty 
reasonable since it is determined primarily by the energies, and those are  
reproduced well by the SV potential. 

The new potential form (SS) introduced by Sahu and Sahu
\cite{[SS12]} can be considered as a generalization of the SV form. 
The extra diffuseness parameter may smooth or roughen the potential 
in the region around $\rho_1$ depending on whether $a_s>1$ or $a_s<1$. 

The range of the SFR potentials determines approximately the starting points 
of the pole trajectories belonging to potential strength zero. 
The problem of the $S$-matrix poles becomes ill-defined in a potential with 
strength $V_0\approx 0$, thus it is important to see whether the computer code 
is able to solve the problem for small $V_0$. A check is provided 
by potentials of the form of 
$-V_0(R-r)$ ($r\le R$),
for which these starting points are approximately determined apart from 
an additive constant. This check has shown that our calculations are 
remarkably accurate. 

It is more surprising that even though the CWS and the SV potentials are very 
different in the neighborhood of the cutoff, the pole trajectories of 
the SV potentials bear out some of the properties of those of 
the 
$-V_0(R-r)$ 
potentials, especially for large node numbers. 
For some low values of the node number, the CWS trajectory shows strange 
shapes, while the SS and SV potentials behave absolutely regularly. 
The pole trajectories of the SS potential depend weakly on the extra 
diffuseness parameter.

In conclusion, the present results are reassuring concerning the use of the 
SFR potentials. The starting points of the pole trajectories seem to have 
some approximate universality properties, which can be used to estimate 
the values of these starting points.

\section*{Acknowledgment}
This work was partially supported by the ENIAC CSI  No. 120209 project and by the  T\'AMOP-4.2.2.C-11/1/KONV-2012-0001 project.
The later project has been supported by the European Union, co-financed by
the European Social Fund.

\end{document}